\documentclass[11pt]{article}
\def\bea{\begin{eqnarray}}
\def\eea{\end{eqnarray}}

\def\beq{\begin{equation}}
\def\eeq{\end{equation}}
\def\ba{\beq\new\begin{array}{c}}
\def\ea{\end{array}\eeq}
\def\be{\ba}
\def\ee{\ea}

\makeatletter
\newdimen\normalarrayskip % skip between lines
\newdimen\minarrayskip % minimal skip between lines
\normalarrayskip\baselineskip
\minarrayskip\jot
\newif\ifold \oldtrue \def\new{\oldfalse}
\def\arraymode{\ifold\relax\else\displaystyle\fi} % mode of array entries
\def\eqnumphantom{\phantom{(\theequation)}} % right phantom in eqnarray
\def\@arrayskip{\ifold\baselineskip\z@\lineskip\z@
\else
\baselineskip\minarrayskip\lineskip2\minarrayskip\fi}
\def\@arrayclassz{\ifcase \@lastchclass \@acolampacol \or
\@ampacol \or \or \or \@addamp \or
\@acolampacol \or \@firstampfalse \@acol \fi
\edef\@preamble{\@preamble
\ifcase \@chnum
\hfil$\relax\arraymode\@sharp$\hfil
\or $\relax\arraymode\@sharp$\hfil
\or \hfil$\relax\arraymode\@sharp$\fi}}
\def\@array[#1]#2{\setbox\@arstrutbox=\hbox{\vrule
height\arraystretch \ht\strutbox
depth\arraystretch \dp\strutbox
width\z@}\@mkpream{#2}\edef\@preamble{\halign
\noexpand\@halignto
\bgroup \tabskip\z@ \@arstrut \@preamble \tabskip\z@ \cr}%
\let\@startpbox\@@startpbox \let\@endpbox\@@endpbox
\if #1t\vtop \else \if#1b\vbox \else \vcenter \fi\fi
\bgroup \let\par\relax
\let\@sharp##\let\protect\relax
\@arrayskip\@preamble}
\def\eqnarray{\stepcounter{equation}%
\let\@currentlabel=\theequation
\global\@eqnswtrue
\global\@eqcnt\z@
\tabskip\@centering
\let\\=\@eqncr
$$%
\halign to \displaywidth\bgroup
\eqnumphantom\@eqnsel\hskip\@centering
$\displaystyle \tabskip\z@ {##}$%
\global\@eqcnt\@ne \hskip 2\arraycolsep
$\displaystyle\arraymode{##}$\hfil
\global\@eqcnt\tw@ \hskip 2\arraycolsep
$\displaystyle\tabskip\z@{##}$\hfil
\tabskip\@centering
&{##}\tabskip\z@\cr}
\begingroup\ifx\undefined\newsymbol \else\def\input#1 {\endgroup}\fi

\def\i{{\mathrm i}}
\def\e{{\mathrm e}}
\def\CF{\bigl\langle\,:\!\bar\phi\ast\phi(x_{(1)})\!:\,
 :\!\bar\phi\ast\phi(x_{(2)})\!:\,\bigr\rangle}
\def\sign{\mathop{\mathrm{sign}}}
\def\tr{\mathop{\mathrm{tr}}}

\begin{document}

\setcounter{footnote}{1}
\def\thefootnote{\fnsymbol{footnote}}
\begin{center}
\hfill ITEP-TH-01/01\\ \hfill hep-th/0101059\\ \vspace{0.3in}
{\Large\bf Correlation Functions of the Scalar Field\\ in
Background NC U(1) Yang-Mills.}
\end{center}
\centerline{{\large A.Solovyov}\footnote{ Kiev University and
BITP, Kiev, Ukraine\\ e-mail: solovyov@phys.univ.kiev.ua}}

\bigskip

\abstract{\footnotesize We consider the complex scalar field
coupled to background NC U(1) YM and calculate the correlator of
two gauge invariant composite operators. We show how the
noncommutative gauge invariance is restored for higher correlators
(though the Green's function itself is not invariant). It is
interesting that the recently discovered noncommutative solitons
appear in the calculation.}

\begin{center}
\rule{5cm}{1pt}
\end{center}

\bigskip
\setcounter{footnote}{0}
\renewcommand\thefootnote{\arabic{footnote}}
\section{Introduction}
Noncommutative gauge theories have attracted attention due to
recent achievements in string theory. Seiberg and Witten in their
celebrated work have shown that the open string theory in large
external constant $B$-field yields the effective noncommutative
theory on the world volume of D-brane\cite{SW}. That is why
coupling of a noncommutative theory to a commutative one can be
useful for understanding the interaction of open and closed
strings.

It is the fundamental idea that the noncommutativity is a
deformation of the algebra of functions, thus the fields of a
noncommutative gauge theory are valued in the deformed algebra of
functions where the usual product is replaced by the associative
noncommutative Moyal $\ast$-product
\be f_1\ast f_2(x)= f_1(x)\e
^{\frac{\i}{2}\overleftarrow{\partial}_\mu\theta^{\mu\nu}
\overrightarrow{\partial}_\nu}f_2(x).
\ee
The following integral
representation of the $\ast$-product will be used in the future,
\be
f_1\ast f_2(x)=\int {d^dx'\,d^dx''\,K(x,x',x'')f_1(x')f_2(x'')},\\
K(x,x',x'')=\frac{1}{\pi^d \det(\theta^{\mu\nu})}
\exp\{-2\i(x'-x)^\mu(\theta^{-1})_{\mu\nu}(x''-x)^\nu \}. \ee We
shall restrict ourselves to the case of two noncommuting spatial
coordinates, $$\left[ x^1,x^2 \right]=\i\theta^{12}\equiv
\i\theta.$$ In this case\cite{Za} \be \label{eq3}
K(x,y,z)=\frac1{\pi^2\theta^2} \e^{-\frac{2\i}{\theta}
\bigl(x^2(y^1-z^1)+y^2(z^1-x^1)+z^2(x^1-y^1) \bigr)}. \ee This
formula will be exploited in the subsequent calculations to
evaluate different $\ast$-products. If some extra commuting
temporal or spatial coordinates are present they serve as
additional parameters from the $\ast$-product view-point.

The Weyl-Moyal (WM) correspondence is a very useful tool for
performing calculations in noncommutative theories. This is the
one-to-one correspondence (isomorphism) between the deformed
algebra of functions (i.e. the former algebra to which
noncommutative fields belong) on a noncommutative manifold with a
constant noncommutativity matrix $\theta^{ij}$ and the algebra of
operators acting in the auxiliary Hilbert space. The literature on
this topic is very extended so we do not spend too much time for
discussing it (for example, see\cite{BFFLS}).

The paper is organized as follows. In the next section we present
the model, i.e{.} its action functional and gauge transformations.
This is the massive charged scalar field $\phi$ coupled to a
background noncommutative gauge field $A$. Further we consider a
particular background of constant strength. The third section
deals with the correlation functions of the theory. Such a free
(quadratic in $\phi$) theory appears to be in some sense
equivalent to an effective commutative one. This equivalence
becomes especially transparent physically when we use the linear
ansatz for the gauge field\cite{A-GB}. Namely, a gauge-dependent
rescaling of coordinates is required to perform this reduction.
That is why the definition of the gauge invariant correlators
proves to involve the $\ast$-product and differs from that in the
commutative case. It is the $\ast$-product that is expected to
rescue the situation. The gauge invariance of these correlators is
verified by the explicit calculation using the formal spectral
definition of the Green's function. These objects are shown to
really possess the required feature. The gauge invariant two-point
function is calculated explicitly as an infinite series where each
term is obviously gauge invariant.
\section{Classical theory}
The model we consider is the complex scalar field theory coupled
to background noncommutative U(1) Yang-Mills. The action
functional is given by \be \label{eq4} S=-\int\nolimits{\bar\phi
(-D_i D^i+m^2)\phi}. \ee The metric tensor is considered to be of
euclidean $(++)$ signature. The covariant derivatives act
according to \be
D_i\phi=\partial_i\phi-\i A_{i}\ast\phi,\\
F_{ij}=\i\left[D_i,D_j\right]=
\partial_i A_j-\partial_j A_i -\i[A_i,A_j].
\ee The gauge transformations \be \phi \rightarrow U\ast\phi,\quad
\bar\phi\rightarrow\bar\phi\ast\bar U,\\ A_i\rightarrow U\ast A_i
\ast \bar U -\i \partial_i U \ast \bar U,\\ F_{ij}\rightarrow
U\ast F_{ij}\ast\bar U \ee are generated by a star-unitary $U$
such that \be \label{eq6} \bar{U}\ast U =1= U \ast\bar{U}. \ee In
what follows we shall work with the background of constant
strength $F_{12}=F$ and choose the potential \be \label{eq8}
A_1=-\alpha_1 x^2,\quad A_2=\alpha_2 x^1,\\ F=
\alpha_1+\alpha_2+\theta\alpha_1\alpha_2. \ee Though $F$ itself is
not gauge invariant, in the present situation it is so. $F>0$ will
be assumed hereinafter without loss of generality. Without
analyzing in detail all the gauges\cite{A-GB} we simply exhibit
the one-parametric family of the functions $U$ that generate some
gauge transformations leaving the potential like (\ref{eq8})
within this class: \be U_t=\frac{1}{\cosh
t}\e^{\frac{2\i}{\theta}x^1x^2\tanh t},\\ U_0=1, \quad
\bar{U}_t=U_{-t},\\ U_{t_1}\ast U_{t_2}=U_{t_1+t_2}. \ee The gauge
transformation generated by $U_t$ changes $\alpha_i$ as \be
\alpha_1\rightarrow \e^{-2t}\alpha_1 -\frac2\theta \e^{-t}\sinh
t,\\ \alpha_2\rightarrow \e^{2t}\alpha_2 +\frac2\theta \e^t \sinh
t. \ee Another interesting thing is that the invariance of the
action (\ref{eq4}) is provided even if \footnote{Author thanks
A.Morozov for pointing at this fact.} \be \label{eq11} \bar U\ast
U=1, \quad U\ast\bar U=1-P. \ee Obviously $P\ast P=P$, i.e. $P$ is
a projector. This generating function is `topologically
nontrivial', i.e. $U\neq \e_\ast^{\i f}$ for any real $f$. Field
strength transforms according to \be F_{ij}\rightarrow U\ast
F_{ij}\ast\bar U +U\ast(A_j\ast\partial_i\bar U -
A_i\ast\partial_j \bar U)\ast P +\i(\partial_iU\ast\partial_j\bar
U -
\partial_jU\ast\partial_i\bar U)\ast P
\ee
but this makes no trouble as the $F_{ij}F^{ij}$ term is not
present in the action for background $A_i$. In the case of such a
transformation the gauge
field $A_i$ no longer remains real.
\section{Quantum theory}
It has been emphasized that free noncommutative theories are
identical to the commutative ones, so the definition of the free
propagator in the background gauge field is the same, i.e. as the
inverse to the quadratic in $\phi$ part of the action. If $f_n$
are orthonormalized eigenfunctions of the operator $(-D_i D^i
+m^2)$ with the corresponding eigenvalues $\lambda_n$ then the
Green's function is given by its formal spectral definition
\footnote{Upper indices of $x$'s are coordinate indices and those
in brackets will onwards mark the number of point in correlator.}
\be \label{eq13}
G(x_{(1)},x_{(2)})=-\sum\limits_{n}{}{\frac{1}{\lambda_n}f_n(x_{(1)})
\bar{f_n}(x_{(2)})}, \ee $n$ being discrete or continuous. As in
the usual case, \be
\left\langle\phi(x_{(1)})\bar\phi(x_{(2)})\right\rangle=
\int\!{\mathcal D}\phi\,{\mathcal D}\bar \phi\, \e^{\i S[\phi,\bar
\phi]}\phi(x_{(1)})\bar\phi(x_{(2)})=\i G(x_{(1)},x_{(2)}). \ee
This free propagator is not gauge invariant and transforms as \be
\label{eq15} G(x_{(1)},x_{(2)})\rightarrow U(x_{(1)})\ast
G(x_{(1)},x_{(2)})\ast \bar U(x_{(2)}). \ee To verify this we have
to check the invariance of the measure w.r.t{.} the gauge
transformations. Now it becomes clear why both conditions of
(\ref{eq6}) are necessary. An example of $U$ satisfying
(\ref{eq11}) and not satisfying (\ref{eq6}) is\cite{HKL} \be \hat
U:|n\rangle\rightarrow |n+1\rangle. \ee (we have used the WM
correspondence). If one now approximates the path integral measure
as \be {\mathcal D}\phi\, {\mathcal
D}\bar\phi= N\prod\limits_{m,n\geq 0}{d\phi_{mn} d\bar\phi_{mn}},\\
\phi_{mn}=\langle m|\hat \phi|n \rangle, \quad
\bar\phi_{mn}=\langle m|\hat{\phi}^\dagger|n \rangle \ee where
$\hat\phi$ denotes the operator to which $\phi$ is mapped under
the Weyl-Moyal correspondence then \be \langle m| \hat U \hat\phi
|n\rangle= \left\{
\begin{array}{lr}
\phi_{m-1,n}, & m\geq 1\\
0, & m=0
\end{array} \right.
\ee The similar formula holds for
$\bar{\phi}\leftrightarrow\hat{\phi}^\dagger$ and even the domain
of integration is not invariant. Another way to obtain the
transformation law (\ref{eq15}) is expansion (\ref{eq13}). If the
transformation generated by $U$ is invertible, i.e. $U$ has a left
inverse that can generate a gauge transformation (in the case of
(\ref{eq11}) $\bar U$ cannot do it) then all the eigenfunctions
before and after the transformation are in one-to-one
correspondence and (\ref{eq15}) becomes obvious.

If one chooses $A_1=-Fx^2,$ $A_2=0,$ then
$D_1=(1+\frac{F\theta}{2})\partial_1+\i Fx^2,$ $D_2=\partial_2$.
Thereby is the effect of noncommutativity merely rescaling the
$x^1$ coordinate? On the other side, the gauge $A_1=0,$ $A_2=Fx^1$
implies rescaling of the other coordinate. Should one use the
symmetric gauge, both coordinates would be rescaled by an
identical factor (this situation seems to respect the rotational
symmetry more than the two former ones). It becomes clear from
this simple example that the subject of interest (in our case
gauge invariant correlators) is also changed w.r.t{.} the
commutative case. So the naive correlator
$\langle\,:\!\bar{\phi}\phi(x_{(1)})\!:\,
:\!\bar{\phi}\phi(x_{(2)})\!:\,\rangle$ no more remains gauge
invariant and should be replaced by \footnote{This correlator will
be referred to as a two-point one.} \be \label{eq19} \CF. \ee Let
us denote $\beta_i=1+\frac{\alpha_i \theta}{2},$ then the
covariant derivatives take the form \be
D_1=\beta_1\partial_1+\i\alpha_1 x^2,\\
D_2=\beta_2\partial_2-\i\alpha_2 x^1. \ee The problem of finding
the above eigenfunctions $f_n$ can be solved using the ansatz
$f_n(x)=\exp(\i\frac{\alpha_2}{\beta_2}x^1x^2)g_n(x)$; it reduces
to \be \left\{-\beta_1^2\partial_1^2-\beta_2^2\partial_2^2
-2\i\bigl(\frac{\beta_1^2\alpha_2}{\beta_2}
+\alpha_1\beta_1\bigr)x^2\partial_1
+\bigl(\frac{\beta_1^2\alpha_2}{\beta_2}
+\alpha_1\beta_1\bigl)^2(x^2)^2+m^2 \right\}g_n =\lambda_n g_n.
\ee From this one easily finds requisite
eigenfunctions/eigenvalues \be
f_{n,k}=\frac{\sqrt[4]{F}}{\sqrt{2\pi|\beta_2|}}
\exp\left\{\i\bigl(\frac{\alpha_2}{\beta_2}x^1x^2+kx^1\bigr)\right\}
\psi_n\left(\frac{x^2\sqrt{F}}{\beta_2}+\frac{\beta_1k}{\sqrt{F}}\right),\\
\lambda_n=(2n+1)F+m^2.
\ee
Here $\psi_n$ stands for the $n$-th
normalized wavefunction of one-dimensional harmonic oscillator
with frequency equal to unity, i.e. $\e
^{-\frac{x^2}{2}}$ multiplied by some Hermite polynomial. To
obtain the correlator $(\ref{eq19})$ one can use the ordinary Wick's
theorem coming from the commutative case arriving to
\be
\label{eq23}
\CF = -G(x_{(1)},x_{(2)})
e^{\frac{\i}{2}\theta^{ij}
(\overrightarrow{\partial}_{(1)i}\overleftarrow{\partial}_{(1)j}
+\overleftarrow{\partial}_{(2)i}\overrightarrow{\partial}_{(2)j})
} G(x_{(2)},x_{(1)}).
\ee
The answer is
\be
\label{eq24}
\CF=\\
-\sum\limits_{n_1,n_2=0}^{\infty}
{\frac{1}{\lambda_{n_1}\lambda_{n_2}}
\int\limits_{-\infty}^{+\infty} \int\limits_{-\infty}^{+\infty}
dk_1\, dk_2 \bigl\{\bar{f}_{n_1,k_1}\ast
f_{n_2,k_2}(x_{(1)})\bigr\} \bigl\{\bar{f}_{n_2,k_2}\ast
f_{n_1,k_1}(x_{(2)})\bigl\} }.
\ee
The integral kernel (\ref{eq3}) is of extreme use
for the evaluation of the rhs
of (\ref{eq24}). The result is
\be
\label{eq25}
\bar{f}_{n_1,k_1}\ast f_{n_2,k_2}(x)=
-\frac{|2+\alpha_2\theta|\sqrt {F}}{4\pi|1+\alpha_2\theta|}
\psi_{n_1} \left(\frac{(2+\alpha_2\theta)(2+F\theta)}
{4(1+\alpha_2\theta)\sqrt{F}}k_1+ \frac{(2+\alpha_2\theta)F\theta}
{4(1+\alpha_2\theta)\sqrt{F})}k_2+x^2\sqrt{F}\right)\\ \psi_{n_2}
\left(\frac{(2+\alpha_2\theta)F\theta}
{4(1+\alpha_2\theta)\sqrt{F})}k_1+
\frac{(2+\alpha_2\theta)(2+F\theta)}
{4(1+\alpha_2\theta)\sqrt{F}}k_2+x^2\sqrt{F}\right) \e
^{\i x^1\frac{(k_2-k_1)(2+\alpha_2\theta)}{2(1+\alpha_2\theta)}}.
\ee
In the future the following substitution will be useful:
\be
\label{eq26}
k_1=\frac{\sqrt{F}}{2+F\theta+\alpha_1\theta}
\bigl((2+F\theta)\xi_1-F\theta\xi_2\bigr),\\
k_2=\frac{\sqrt{F}}{2+F\theta+\alpha_1\theta}
\bigl(-F\theta\xi_1+(2+F\theta)\xi_2\bigr),\\ \Bigl|\,
\det\Bigl(\frac{\partial(k_1,k_2)}{\partial(\xi_1,\xi_2)}\Bigr)\,\Bigr|=
\frac{4F(1+\alpha_2\theta)}{(1+\alpha_1\theta)(2+\alpha_2\theta)^2}
\sign(1+F\theta) \ee as then \be \bar{f}_{n_1,k_1}\ast
f_{n_2,k_2}(x)= -\frac{|2+\alpha_2\theta|\sqrt{F}}{4\pi
|1+\alpha_2\theta|} \psi_{n_1}\left(x^2\sqrt{F}+\xi_1\right)
\psi_{n_2}\left(x^2\sqrt{F}+\xi_2\right)
\e ^{\i
x^1\sqrt{F}(\xi_2-\xi_1)}.
\ee
As $\lambda_n$'s do not depend on
$k$, the integration over $k_1,k_2$ (or, equivalently, $\xi_1,\xi_2$) in
(\ref{eq24}) can be done explicitly and the result is proportional
to (we have performed the constant shift in the integration variables
$\xi_i\rightarrow\xi_i-\frac{(x_{(1)}^2+x_{(2)}^2)\sqrt{F}}{2}$)
\be
\int d\xi_1\, d\xi_2\,
\psi_{n_1}\left(\frac{x^2\sqrt{F}}{2}+\xi_1\right)
\psi_{n_1}\left(-\frac{x^2\sqrt{F}}{2}+\xi_1\right)\\
\times\psi_{n_2}\left(\frac{x^2\sqrt{F}}{2}+\xi_2\right)
\psi_{n_2}\left(-\frac{x^2\sqrt{F}}{2}+\xi_2\right)
\exp{\left\{2\i\frac{x^1\sqrt{F}}{2}(\xi_1-\xi_2)\right\}},\\
x\equiv x_1-x_2.
\ee
The crucial feature is the definite parity of
$\phi_n$'s as now
\be
\int d\xi\,
\psi_n\left(-\frac{x^2\sqrt{F}}{2}+\xi\right)
\psi_n\left(\frac{x^2\sqrt{F}}{2}+\xi\right)
\exp\left\{2\i\frac{x^1\sqrt{F}}{2}\xi\right\}=\\ (-1)^n\int
d\xi\, \psi_n\left(\frac{x^2\sqrt{F}}{2}-\xi\right)
\psi_n\left(\frac{x^2\sqrt{F}}{2}+\xi\right)
\exp\left\{2\i\frac{x^1\sqrt{F}}{2}\xi\right\}=
\frac{(-1)^n}{2}\phi_n\left(\frac{x^2\sqrt{F}}{2},
\frac{x^1\sqrt{F}}{2}\right).
\ee
$\phi_n$ denotes the phase space
Wigner function corresponding to the quantum mechanical state
described by $|\psi_n\rangle$, i.e{.} the function to which the
$|\psi_n\rangle \langle\psi_n|$ operator is mapped under the
Weyl-Moyal correspondence ($\hbar=1$). In our case
\be
\phi_n(x)=2
(-1)^n \e^{-|x|^2} L_n(2|x|^2),
\ee
$L_n$ being the $n$-th Laguerre
polynomial. These functions form the complete set of
one-dimensional radially symmetric projectors solving the equation
$\phi\ast\phi=\phi$\cite{GMS}. The final answer reads
\be
\label{eq31}
\CF= -\frac{1}{|1+F\theta|\pi^2}
\left(\sum\limits_{n=0}^{\infty}
{\frac{(-1)^nF\phi_n(\frac{x\sqrt{F}}{2})}{4((2n+1)F+m^2)}}\right)^2,\\
x\equiv x_{(1)}-x_{(2)}. \ee In these calculations the two-point
correlation function factors in a natural way just as in the
commutative theory $\langle\,:\!\bar\phi\phi(x_{(1)})\!:\,
:\!\bar\phi\phi(x_{(2)})\!:\,\rangle
=-|\langle\bar\phi(x_{(1)})\phi(x_{(2)})\rangle|^2$. So in the
noncommutative case (\ref{eq31}) is also a full square (not just a
$\ast$-square) of a gauge invariant quantity.\\ As for $F\to 0 \;
\sum\limits_{n}^{}
\frac{(-1)^nF\phi_n(\frac{x\sqrt{F}}{2})}{4((2n+1)F+m^2)} \sim
\sum\limits_{n}^{}
\frac{(-1)^nF\phi_n(\frac{x\sqrt{F}}{2})}{4m^2}=
\frac{\delta^{(2)}(x)}{m^2}$ the correlator displays singular
behaviour in this limit.

The higher correlators are calculated in the way similar to that of
the commutative
theory with the novel feature of multiplying Green's functions with
the $\ast$-product like (\ref{eq23}) and it is the $\ast$-product that
does provide the
noncommutative gauge invariance. For the evaluation of the
$n$-point function one can rescale the integration variables so
that
\be
\frac{2+\alpha_2\theta}{1+\alpha_2\theta}k_i\rightarrow k_i,
\ee
then the Jacobian cancels the non-invariant factor
coming from the rhs of (\ref{eq25}) and every term of the series
is explicitly gauge
invariant (i.e. expressed in terms of $F$). Correlators with $n>2$ points
do not reduce to
projector solitons of\cite{GMS} anymore, e.g{.} for $n=3$ there appear
terms like
\be
\int dk_1\, dk_2\, dk_3\, \e^ {
\frac \i2 \bigl( x^1_{(1)}(k_3-k_1) +x^1_{(2)}(k_1-k_2)
+x^1_{(3)}(k_2-k_3) \bigr) }\\
\times\psi_{n_1}\left(x^2_{(1)}\sqrt{F}+\frac{(2+F\theta)k_1+F\theta
k_3 }{4\sqrt F}\right)
\psi_{n_3}\left(x^2_{(1)}\sqrt{F}+\frac{(2+F\theta)k_3+F\theta k_1
}{4\sqrt F}\right)\\
\times\psi_{n_2}\left(x^2_{(2)}\sqrt{F}+\frac{(2+F\theta)k_2+F\theta
k_1 }{4\sqrt F}\right)
\psi_{n_1}\left(x^2_{(2)}\sqrt{F}+\frac{(2+F\theta)k_1+F\theta k_2
}{4\sqrt F}\right)\\
\times\psi_{n_3}\left(x^2_{(3)}\sqrt{F}+\frac{(2+F\theta)k_3+F\theta
k_2 }{4\sqrt F}\right)
\psi_{n_2}\left(x^2_{(3)}\sqrt{F}+\frac{(2+F\theta)k_2+F\theta k_3
}{4\sqrt F}\right)
\ee
and there is no use trying to do a substitution like
(\ref{eq26}) unless $\theta=0$. It is easy to see that the latter expression
does not vary when the identical shift in $x_{(i)}$'s is done
so it depends only on the relative position of the points.

It is also useful to construct the generating functional. To do
this we add to the action the current corresponding to the
composite operator $\bar\phi\ast\phi$:
\be
S\rightarrow S+\int
J(x)\,\bar\phi\ast\phi(x) =S+\int
A(x',x'')\bar\phi(x')\phi(x''),\\ A(x',x'')=\int dx\,
J(x)K(x,x',x'').
\ee
Then
\be
Z[J]=N\det(\i G^{-1}+\i
A)=\det(1+GA).
\ee
The generating functional for connected
diagrams
\be
W[J]=\log Z[J]=\tr\log (1+GA)=\\
\tr(GA)-\frac12\tr(GA)^2+\ldots
\ee
Normal ordering
$:\!\bar\phi\ast\phi\!:$ is nothing but removing the first term from
the rhs manually. Obviously the former results are recovered because
the variation $\frac{\delta A(x',x'')}{\delta J(x)}$ produces
$K(x,x',x'')$ so after the integration over $x'$ and $x''$
the $\ast$-product is reproduced.
\section{Concluding remarks}
All the previous results can be easily generalized to the more realistic
case of
2+1-dimensional field theory in the constant magnetic field. The
Green's functions are
\be
G(x_{(1)},x_{(2)})=-\frac \i2\sum\limits_n
\frac{\e^{-\i\sqrt{\lambda_n}|x^0_{(1)}-x^0_{(2)}|}}{\sqrt \lambda_n}
f_n(\vec x_{(1)}) \bar f_n(\vec x_{(2)})
\ee
with the same $f_n$'s. So the most interesting features survive.
The 2-point gauge invariant functions can still be expressed in terms of the
Wigner functions (noncommutative projector solitons). This
statement seems to be valid for a large class of potentials.

The gauge invariance is recovered with the help of the
$\ast$-product between the Green's functions that replaces the usual
one and our result does depend on coordinates with \emph{no}
gauge-dependent rescaling  resolving the seeming paradox
(what naively does not look completely obvious).

The main goal of the paper is to verify the above physically
nontrivial statements concerning noncommutative gauge invariance
etc. explicitly.
\section{Acknowledgements}
Author is grateful to ITEP group for hospitality and numerous
discussions, in particular to K.Saraikin, A.Gorsky, K.Selivanov
and A.Morozov for useful discussions and carefully reading the
manuscript and especially to A.Morozov for initiating this work.
Author thanks Yu.Sitenko and V.Shadura for support and
discussions. The work was supported by grant INTAS-99-590.

\end{document}